\documentclass[a4paper,11pt]{article}
\usepackage{pos}
\usepackage{wrapfig}

\DeclareMathOperator{\Tr}{Tr}

\def\beann{\begin{eqnarray*}} 
\def\eeann{\end{eqnarray*}}
\def\D{{\cal D}}

\def\T{{\cal T}}
\def\M{{\cal M}}
\def\P{{\cal P}}

\def\I{{\mathbb{I}}}

\newcommand{\Bern}{Albert Einstein Center for Fundamental Physics, Institute for Theoretical Physics, University of Bern, Sidlerstrasse 5, CH--3012 Bern, Switzerland
}
\newcommand{\CERN}{CERN, Theoretical Physics Department, CH-1211 Genève 23, Switzerland}

\title{Heavy-dense QCD at fixed baryon number without a sign problem}

\ShortTitle{Heavy-dense QCD at fixed baryon number without a sign
  problem}

\author[a]{Patrick B\"uhlmann}\affiliation[a]{\Bern}
\author*[a,b]{Urs Wenger}\affiliation[b]{\CERN}

\emailAdd{ventura@itp.unibe.ch}
\emailAdd{wenger@itp.unibe.ch}

\abstract{QCD at fixed baryon number can be formulated in terms of
  transfer matrices explicitly defined in the canonical sectors.  In
  the heavy-dense limit, the fermionic contributions to the canonical
  partition functions in terms of Polyakov loops and quark occupation
  numbers turn out to be completely factorized in space. At low
  temperatures and infinitely strong coupling the sign problem is
  reduced by orders of magnitude for any baryon number as compared to
  the corresponding grand-canonical ensemble. In the canonical
  formulation it is straighforward to integrate out the Polyakov loops
  in the fermionic weights yielding the partition function as a sum of
  only baryon occupation numbers in which the sign problem is
  absent. Using an effective form of the gauge action valid for small
  values of the gauge coupling, the same can be achieved away from the
  strong coupling limit in terms of quark occupation numbers and
  fluxes which couple the quarks with each other. The emerging
  clusters suggest the construction of algorithms which circumvent the
  sign problem in the heavy-dense limit including the full gauge
  action for any value of the gauge coupling.  }

\FullConference{%
 The 38th International Symposium on Lattice Field Theory, LATTICE2021
  26th-30th July, 2021
  Zoom/Gather@Massachusetts Institute of Technology
}

\begin{document}
\maketitle

\section{Motivation}
Let us briefly summarize our motivation for investigating the
canonical formulation instead of the commonly used grand-canonical
one. Consider the grand-canonical partition function in terms of
the Hamiltonian ${\cal H}$ at finite
chemical potential $\mu$ and temperature $T$,
\[
Z_\text{GC}(\mu) = \Tr \left[e^{-{\cal H}(\mu)/T}\right] = \Tr \prod_t
\T_t(\mu) \, .
\]
The calculation of the trace may suffer from a sign problem depending
on the choice of the basis states over which the trace is
taken. The sign problem manifests itself in cancellations between
different states: while all states are present for any values of $\mu$
and $T$, different states need to cancel out for different values of
$\mu$ and $T$. In QCD, for example, at high temperatures the states
describing deconfined quarks are highly relevant and provide the
dominant contributions to the partition function, while in the
confined phase at low temperatures those contributions need to cancel
out. In the canonical formulation, the partition function reads
\[
Z_\text{C}(N_q) = \Tr_{N_q}\big[ e^{-{\cal H}/T}\big]  = \Tr_{N_q} \T=
\Tr \prod_t \T^{(N_q)}_t \, ,
\]
where the sum is now restricted to the states with a fixed number of
fermions or quarks $N_q$. Therefore, it is clear that the dimension of
the Fock space is tremendously reduced and much less cancellations are
necessary. As an example, consider again QCD where in the canonical
formulation it is explicit that $Z^{SU(N_c)}_\text{C}(N_q) = 0$ for
$N_q \neq 0$ mod $N_c$ with the obvious implications, while in the
grand-canonical formulation the corresponding physical principle is
achieved only through implicit cancellations of unphysical states
involving quark numbers $N_q \neq 0$ mod $N_c$. Another example
concerns the Schwinger model where it is explicit that
$Z^{U(1)}_\text{C}(N_q) = 0$ for $N_q \neq 0$, i.e., only noncharged
states are physical, while in the grand-canonical ensemble the same
physical property emerges only after the cancellation of the
contributions from all charged states. Finally, we point out that the
so-called "Silver Blaze" phenomenon is realised automatically in the
canonical formulation, in contrast to the grand-canonical one.
              
In the expressions above we have indicated that the trace is taken
over a product of transfer matrices $\T=\prod_t \T_t$. The
conservation of fermion number manifests itself in the fact that the
transfer matrices are block diagonal where the block matrices are the
canonical transfer matrices $\T_t^{(N_q)}$. It turns out that these
transfer matrices and their product are known in closed form and can
be calculated explicitly in terms of the dimensionally reduced
matrices \cite{Steinhauer:2014oda}. The connection between the
canonical and grand-canonical ensemble is given by the fugacity
expansion
\[
\det M[{\cal U};\mu] = \sum_{{N}_q} e^{-{N_q} \,\mu/T } \cdot  \det{}_{N_q} M[{\cal U}]
\]
where $M[{\cal U};\mu]$ denotes the fermion matrix depending on the
gauge field $U$ and the chemical potential $\mu$. The expansion
relates the grand-canonical fermion determinant to the canonical
ones. Each canonical determinant $ \det{}_{N_q} M[{\cal U}]$ can be
expressed as the trace over the minor matrix ${\cal M}_{N_q}$ of order
${N}_q$ of the dimensionally reduced transfer matrix ${\cal T}$ at
zero chemical potential
\[
\det{}_{N_q} M[{\cal U}]= \sum_{|J|=N_q} \det {\cal T}^{\, \backslash
  \hspace{-0.15cm} J  \, \backslash  \hspace{-0.15cm}J}[{\cal U}]  =
\Tr \prod_t {\cal T}_t^{({N}_q)} = \Tr {\cal M}_{N_q} \, .
\] 
While this connection is completely general, in the following we apply
it to QCD with Wilson fermions in the heavy-dense limit.

\section{Heavy-dense limit of QCD}
The heavy-dense approximation of QCD consists in general of taking the
hopping parameter of the Wilson fermion
$\kappa \equiv (2m+8)^{-1} \rightarrow 0$ and the chemical potential
$\mu \rightarrow \infty$, while keeping the combination
$\kappa e^{+\mu}$ fixed \cite{Blum:1995cb,Engels:1999tz}. This
procedure has the property that only the static quarks are kept as
degrees of freedom, while the static antiquarks are removed from the
system. Another slightly different approximation consists in
distinguishing the spatial and temporal hopping of the fermions,
dropping only the spatial hopping terms and keeping the forward and
backward hopping in time. This procedure has the advantage that one
retains a relativistic system of static quarks {\it and} antiquarks
which is closer to real-world QCD.

In the grand-canonical formulation, the heavy-dense limit, as
described above, leads to a three-dimensional fermion action in terms
of Polyakov loops $P$ and anti-Polyakov loops $P^\dagger$ with the
fermion determinant
\begin{equation}
\det M_{GC}^{HD} = \prod_{\bar x} \det\left[\mathbb{I} - (2\kappa
  e^{+\mu})^{N_t} P_{\bar x}\right]^2 \, \det\left[\mathbb{I} -
  (2\kappa e^{-\mu})^{N_t} P_{\bar x}^\dagger\right]^2 \, ,
\label{eq:detM_GC}
\end{equation}
where $N_t$ is the number of time slices and $\bar x$ denotes the
spatial lattice sites.  As described in the previous section, in the
canonical ensemble the canonical determinants are given by the trace
over the minor matrix $\M_{N_q}$,
\[
\det{}_{N_q} M^{HD} = (2\kappa)^{2 N_c L_s^3 N_t} \cdot  \Tr
  \M_{N_q} \left[ \left( (2\kappa)^{+N_t} \cdot P_+ \P +
      (2\kappa)^{-N_t} \cdot P_- \P \right) \right] \, ,
\]
where $P_\pm=1/2 (\I \mp \gamma_4)$ are Dirac projectors and $\P$
denotes the collection of Polyakov loops $P_{\bar x}$ in the matrix
$\P_{\bar x,\bar y} = \I_{4\times 4}\otimes P_{\bar x} \cdot
\delta_{\bar x,\bar y}$ taking into account the degeneracy with
respect to the Dirac structure. The matrix in square brackets has a
simple block diagonal structure and essentially corresponds to the
dimensionally reduced product of transfer matrices $\T = \prod_t \T_t$
before the explicit projection onto the canonical sectors. This latter
step is achieved by constructing the minor matrix $\M_{N_q}$.

From the structure of the minor matrix ${\cal M}_{N_q}$, or rather the
matrix $\T$ in the square bracket, it is immediately clear that the
canonical determinants transform as
$\det_{N_q} M^{HD} \; \rightarrow \; z_k^{N_q} \cdot \det_{N_q}
M^{HD}$ under a global $z_k \in \mathbb{Z}_{N_c}$
transformation. Consequently, one finds that
\[
  \det{}_{N_q} M = 0 \;\; \text{for} \; N_q\neq 0\mod N_c
\]
which reflects the explicit and exact cancellation of the
contributions from many states in the canonical setup, which is not
evident in the grand-canonical one.

Instead of providing explicit expressions for the canonical
determinants, we just point out that they are simple polynomials of
$\Tr P_{\bar x}$ and $\Tr P^\dagger_{\bar x}$ where all terms have
triality $N_q\mod N_c$ with respect to a $\mathbb{Z}_{N_c}$
transformation. For example, for the canonical determinants with
$N_q=0$ (and $N_c=3$) one finds terms proportional to
$\Tr P_{\bar x}^\dagger\Tr P_{\bar y}$ and
$\Tr P_{\bar x} \Tr P_{\bar y} \Tr P_{\bar z}$, and so on, which are
invariant under global $\mathbb{Z}_3$ transformations, as discussed
above. The terms can be interpreted as contributions from the
propagation of quarks and antiquarks forming mesons and baryons. As a
consequence, the contributions can be described in terms of quark and
antiquark occupation numbers
${n_{\bar x}} \in \{-2 N_c, \ldots, 2 N_c\}$.

It turns out that the system suffers from a severe sign problem,
which, however, is avoided in two cases. Firstly, if all Polyakov
loops $P_{\bar x}$ align along one of the three center elements, which
is the case in the deconfined phase, the contributions from all
Polyakov loops add up coherently and yield a positive
contribution. Secondly, if the global $\mathbb{Z}_{N_c}$ symmetry can
be promoted to a local one, as is the case in the strong coupling
limit, only those contributions survive that satisfy the triality
condition locally, i.e., $n_{\bar x} = 0\mod N_c$.

\section{Heavy-dense strong coupling limit of QCD}
As mentioned above, in the strong coupling limit the global 
$\mathbb{Z}_{N_c}$ transformations are promoted to local ones in the
canonical formulation. This means that since the canonical fermion
determinants transform covariantly under a $\mathbb{Z}_{N_c}$
transformation, integrating locally over all
$\mathbb{Z}_{N_c}$-transformed gauge fields makes all of them vanish
except when $n_{\bar x} = 0\mod N_c$, i.e., the remaining,
nonvanishing contributions are invariant under a local
$\mathbb{Z}_{N_c}$ transformation. This is in contrast to the
grand-canonical setup, where the fermionic contribution
Eq.~(\ref{eq:detM_GC}) does not transform covariantly and is not
$\mathbb{Z}_{N_c}$-invariant in the strong coupling limit. As a
consequence, the partition function becomes a summation over all
baryon configurations $\{n_B(\bar x)\}$ with (essentially) positive
contributions. More precisely, the partition function reads
\begin{equation}
Z_C(N_B) = \left(2 \kappa \right)^{2 N_c N_t L_s^3} \cdot
\sum_{\{n_B\},|n_B|=N_B} \int \D U \prod_{\bar x} \det
\M^{HDSS}_{n_B(\bar x)}[\Tr P_{\bar x}] \, ,
\label{eq:canonicalZHDSS}
\end{equation}
where the fermionic weight is completely factorised in space and
$\det \M^{HDSS}_{n_B(\bar x)}[\Tr P_{\bar x}]$ are the single-site
fermion weights. They are simple polynomials of $\Tr P_{\bar x}$ and
$\Tr P^\dagger_{\bar x}$ with coefficients depending on
$(2\kappa)^{N_t} = \exp(-m/T)$ which parameterizes the temperature.

\begin{wrapfigure}[]{r}{0.5\textwidth}
  \vspace{-0.5cm}
\includegraphics[width=0.5\textwidth]{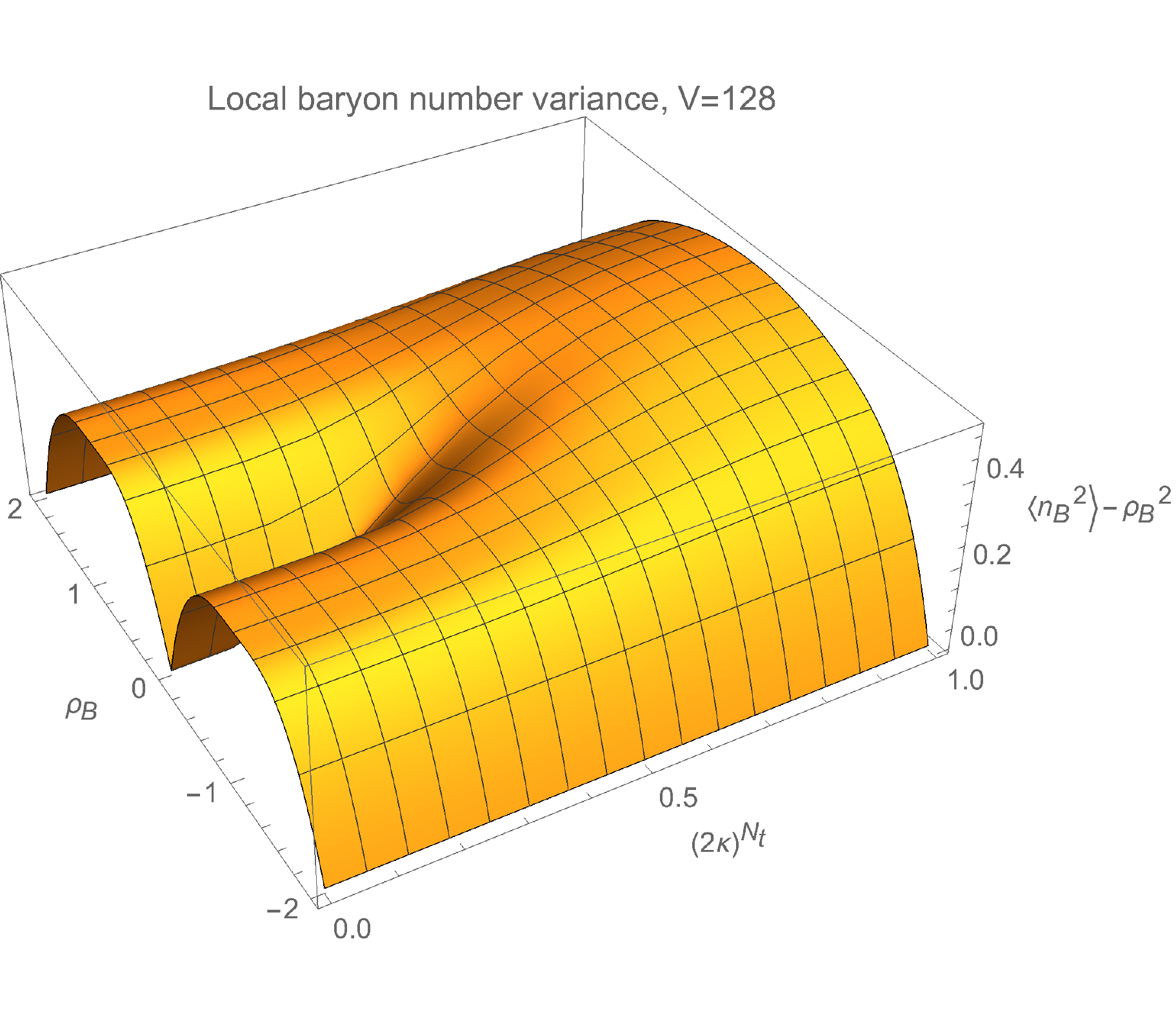}
\caption{Local baryon number susceptibility as a function of the
  baryon density $\rho_B$ and the temperature expressed as
  $\exp(-m/T)=(2\kappa)^{N_t}$. \label{fig:localBaryonNumberVariance}}
\end{wrapfigure}
It is now very easy to analytically integrate out the remaining gauge
field degrees of freedom. This leads to a system where the local baryon
occupation numbers are the only degrees of freedom and where all weights
are positive, i.e., the sign problem at fixed baryon number is solved
in the strong coupling limit. For small system sizes the remaining
summation over the baryon occupation numbers can be done analytically,
while for larger systems one needs to resort to a stochastic
summation. However, it turns out that in the strong coupling limit,
the finite size effects are rather mild and observables typically
saturate already on small systems.

As an example of such a calculation we show in
Figure~\ref{fig:localBaryonNumberVariance} the local baryon number
fluctuations $\chi_{n_B} = \langle n_B^2\rangle - \rho_B^2$, where
$\rho_B=N_B/V$ is the baryon density, as a function of $\rho_B$ and
$\exp(-m/T)=(2\kappa)^{N_t}$. It is clear that the fluctuations are
symmetric w.r.t.~an interchange of baryons and antibaryons and it
would suffice to show only the positive density part. At zero density
and low temperatures, the fluctuations are essentially zero, but
cross over to larger values with growing temperature. At zero
temperature one finds a steep increase of $\chi_{n_B}$ with growing
baryon density, reaching a plateau around half-filling. When the
density is further increased towards saturation, the fluctuations tend
to zero regardless of the temperature.

\begin{figure}
  \includegraphics[width=0.45\textwidth]{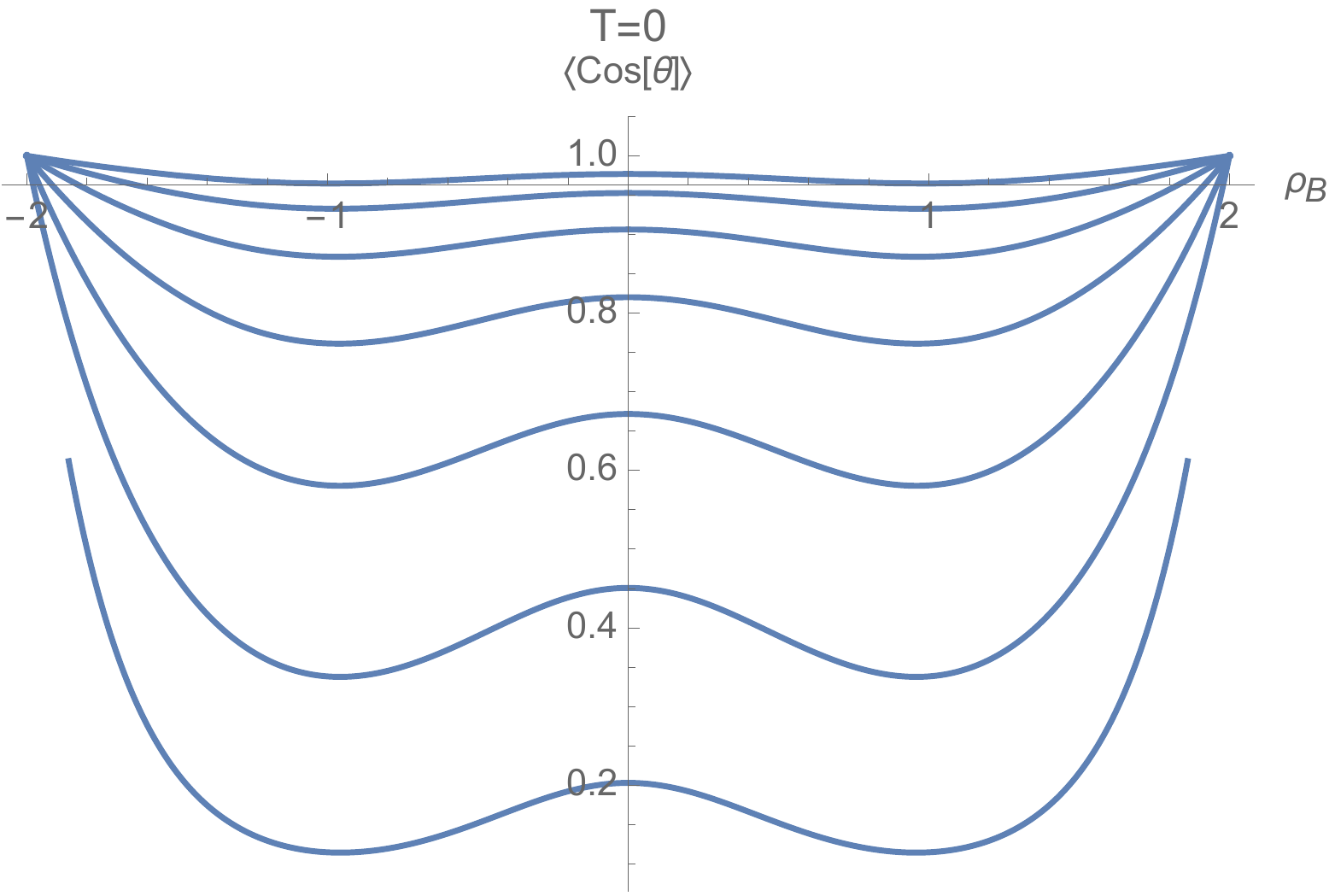}
  \includegraphics[width=0.6\textwidth]{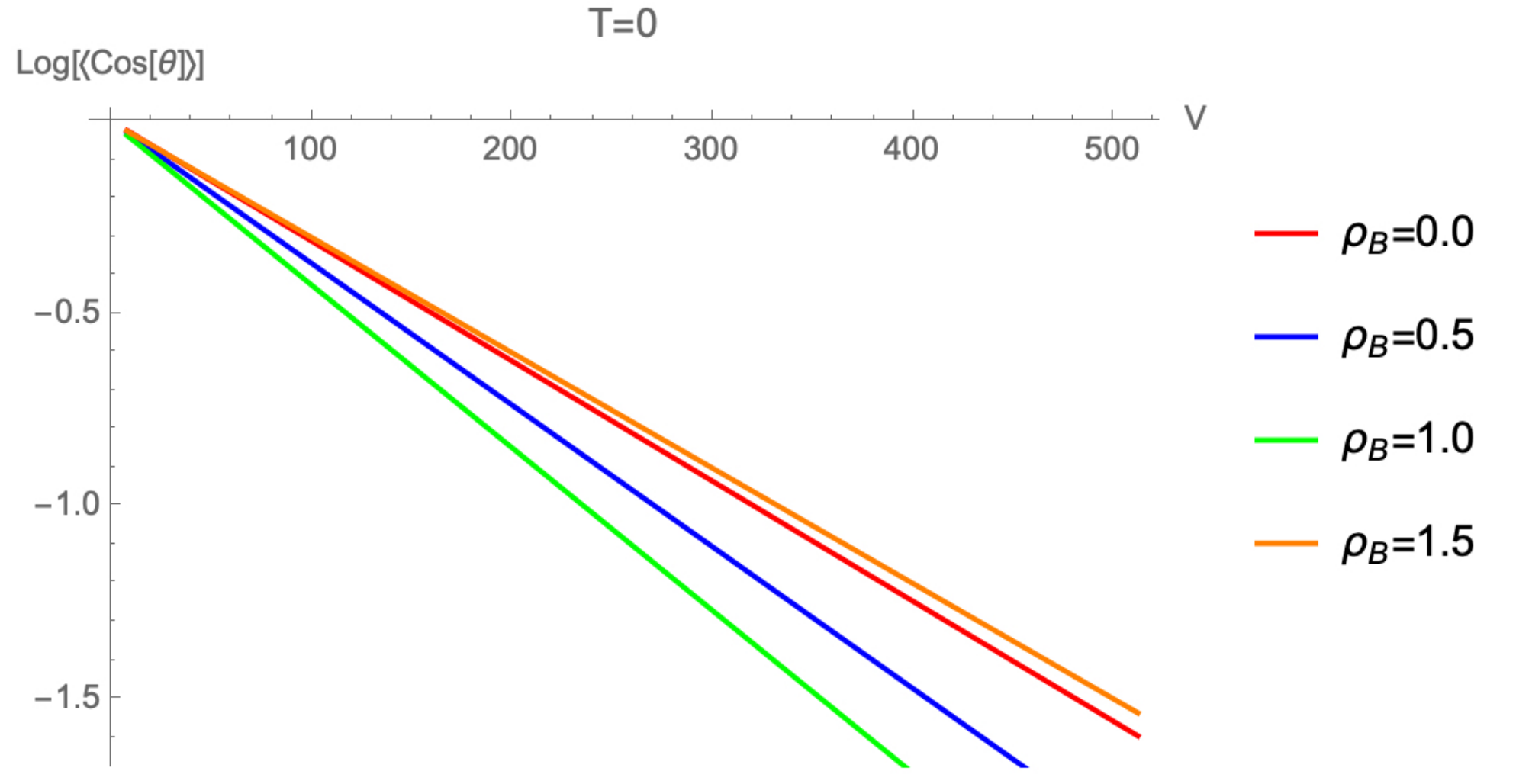}
\caption{{\it Left:} Average sign as a function of the baryon density
  $\rho_B$ for a range of system sizes. {\it Right:}
  Logarithm of the average sign as a function of the system size $V$
  for various baryon densities.\label{fig:V-dependence sign problem}}
\end{figure}

\begin{figure}
  \includegraphics[width=0.6\textwidth]{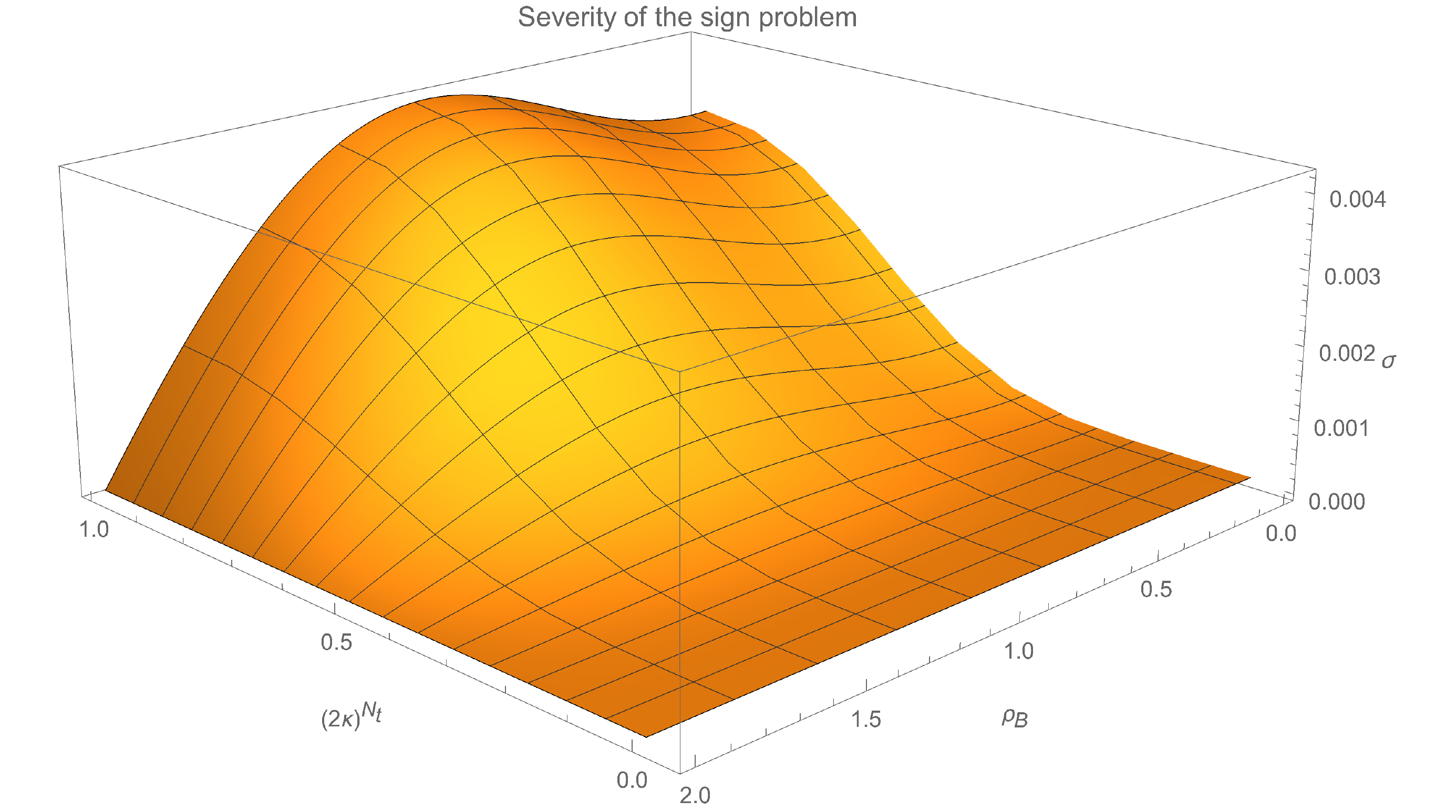} \hfill
  \includegraphics[width=0.425\textwidth]{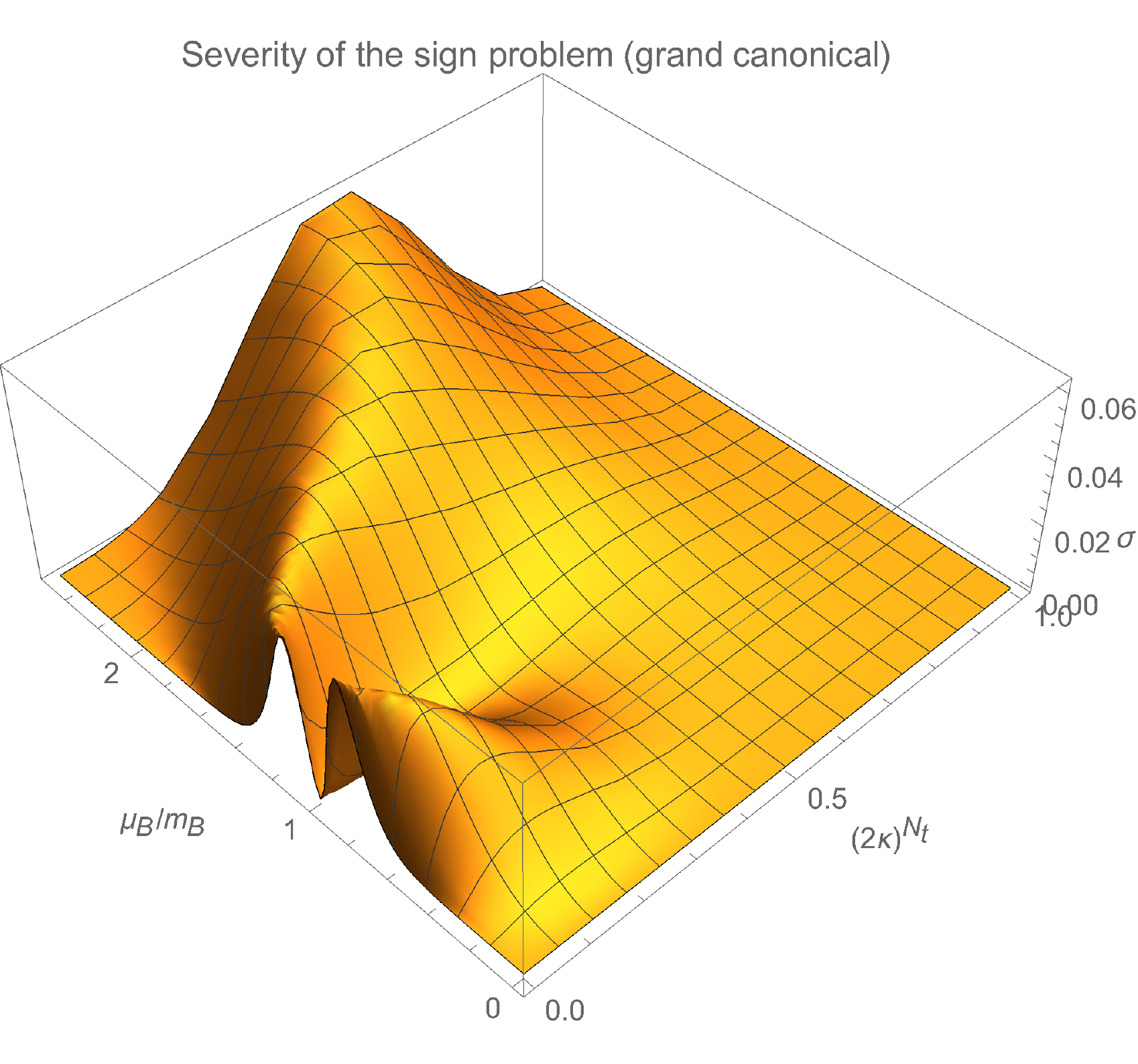}
\caption{Severity of the sign problem as a function of temperature and
  baryon density or baryon chemical potential. {\it Left:} Canonical
  ensemble. {\it Right:} Grand-canonical ensemble.
  \label{fig:anatomy of the sign problem}}
\end{figure}

Of course it is also possible (and instructive) to simulate the
combined system of baryon occupation numbers and the gauge fields,
i.e., the system in Eq.~(\ref{eq:canonicalZHDSS}), in order to assess
the severity of the sign problem. In Figure \ref{fig:V-dependence sign
  problem} we show the average sign
$Z_C(N_B)_{|.|}/Z_C(N_B) = \langle \cos \theta \rangle_{|.|}$ measured
in a phase-quenched simulation. The left plot shows the average sign
as a function of $\rho_B$ for a range of volumes at zero
temperature. While for small system sizes the sign problem is barely
noticable, it becomes more severe with the growing size of the
system. The right plot in Figure \ref{fig:V-dependence sign problem}
shows the logarithm of the average sign as a function of the volume
$V$ for a few selected baryon densities, still at zero temperature.
The exponential decay of the average sign with growing volume
indicates that the sign problem is indeed severe. The severity can be
quantified by determining the free energy density difference between
the phase quenched and the full canonical partition functions,
\begin{equation}
    Z_C(N_B)_{|.|}/Z_C(N_B) = \exp[- \sigma \cdot V] \quad \text{with}
    \quad \sigma = \Delta f/T \, .
\end{equation}
The full anatomy of the sign problem can be uncovered by repeating
this exercise for all temperatures and baryon densities yielding the
left plot in Figure \ref{fig:anatomy of the sign problem}. The sign
problem is most severe at high temperature and half-filling. This can
be understood from the fact that the single-site weight with $n_q=3$
is the only one that can give negative contributions. However, even
there the sign problem is so mild that simulations on rather large
lattices are still practical.
It is interesting to compare this situation to the one in the
grand-canonical ensemble shown in the right plot of Figure
\ref{fig:anatomy of the sign problem} which shows a rather different
behaviour. In particular, at low temperatures the sign problem quickly
becomes severe for increasing $\mu_B$, but at $\mu_B/m_B$, which
corresponds to half-filling $\rho_B = 1$, it vanishes --- a curious
observation that has already been made before
\cite{Rindlisbacher:2015pea}. The less favourable behaviour in the
grand-canonical ensemble (note the different scale on the
$\sigma$-axis) suggests that at finite density (except maybe at
half-filling) and especially at low temperatures, it is preferable to
perform simulations in the canonical ensemble.

\section{Heavy-dense limit of QCD beyond strong coupling}
Trying to go beyond strong coupling and just naively switching on the
gauge coupling ends in complete desaster. For any nonzero gauge
coupling, however small, the sign problem becomes so severe that Monte
Carlo simulations are no longer practical, not even on small
systems. This is due to the fact that as soon as the gauge coupling is
finite, quarks and antiquarks are no longer confined into mesons and
baryons localized on a single site. Instead, they can move away from
each other generating weights with nonzero triality which are no
longer guaranteed to vanish by the local $\mathbb{Z}_{N_c}$
symmetry. In order to get the local $\mathbb{Z}_{N_c}$ symmetry back
into action, we employ the effective gauge action introduced in
\cite{Fromm:2011qi},
 \[
      \exp\left(-S_\text{eff}[{\cal U}]\right) = \prod_{<\bar x \bar
        y>}\left( 1 + 2 \, \lambda \, \text{Re} \Tr P_{\bar x} \Tr P_{\bar
          y}^\dagger\right) \,.
\]
It introduces a nearest-neighbour interaction between Polyakov loops
and can be derived in a systematic way from the original plaquette
gauge action in the heavy-dense limit. As a consequence, the effective
gauge coupling $\lambda$ is in fact a function of the original gauge
coupling and the hopping parameter $\kappa$, and for small $\lambda$
the effective action provides a good description of the full
heavy-dense limit of QCD close to the strong coupling
limit.

On a practical level, since the canonical fermion weights are
factorized in space, it is straightforward to ``dualize'' the
effective nearest-neighbour interaction by introducing fluxes on the
bonds connecting two neighbouring sites. The flux can either be $0$ or
$\pm 1$ and when present on a bond, it induces an additional weight
factor $\lambda$ and additional (anti-)Polyakov loops $P$ and
$P^\dagger$ at the ends of the bond which modify the corresponding
fermion site weights. Integrating the gauge fields locally over the
elements of $\mathbb{Z}_{3}$ induces a local constraint which requires
that the local net flux plus the number of quarks on that site is zero
modulo $3$. That is, quarks and antiquarks, which act as sources and
sinks for the $\mathbb{Z}_{3}$ flux, can now separate from each other,
as long as they are connected by a corresponding flux. This is
essentially a realization of the flux model proposed and elaborated in
\cite{Patel:1983sc,Condella:1999bk}, now derived in a controlled way
from the underlying theory of QCD with Wilson fermions. Finally, the
gauge fields can now be integrated out analytically yielding a system
which is described in terms of integer quark occupation numbers
$n_{\bar x} \in [-6,\ldots, 6]$ and bond occupation numbers
$n_b\in[-1,0,+1]$ with weights that are positive. Hence, the sign
problem is solved beyond strong coupling and Monte Carlo simulations
respecting the local constraints on the integer occupation numbers are
straightforward.

In the following we present a few results as examples of what can be
calculated in this setup. In Figure \ref{fig:(anti-)baryon vs rho} we
show the expectation value of a static baryon or antibaryon in the
background of increasing finite baryon density. Interpreting the
expectation value in terms of the free energy for an additional baryon
or antibaryon, the results demonstrate that it is slightly more
favourable to add an antibaryon to a system with finite baryon density
than it is for a baryon. This is simply due to the fact that an
antibaryon can be screened more easily by the already present baryons
and can therefore more easily be accommodated. Maybe more surprising
is the fact that this situation is reversed once the baryon density
goes beyond half filling $\rho_B > 1$.
\begin{figure}
\includegraphics[width=0.49\textwidth]{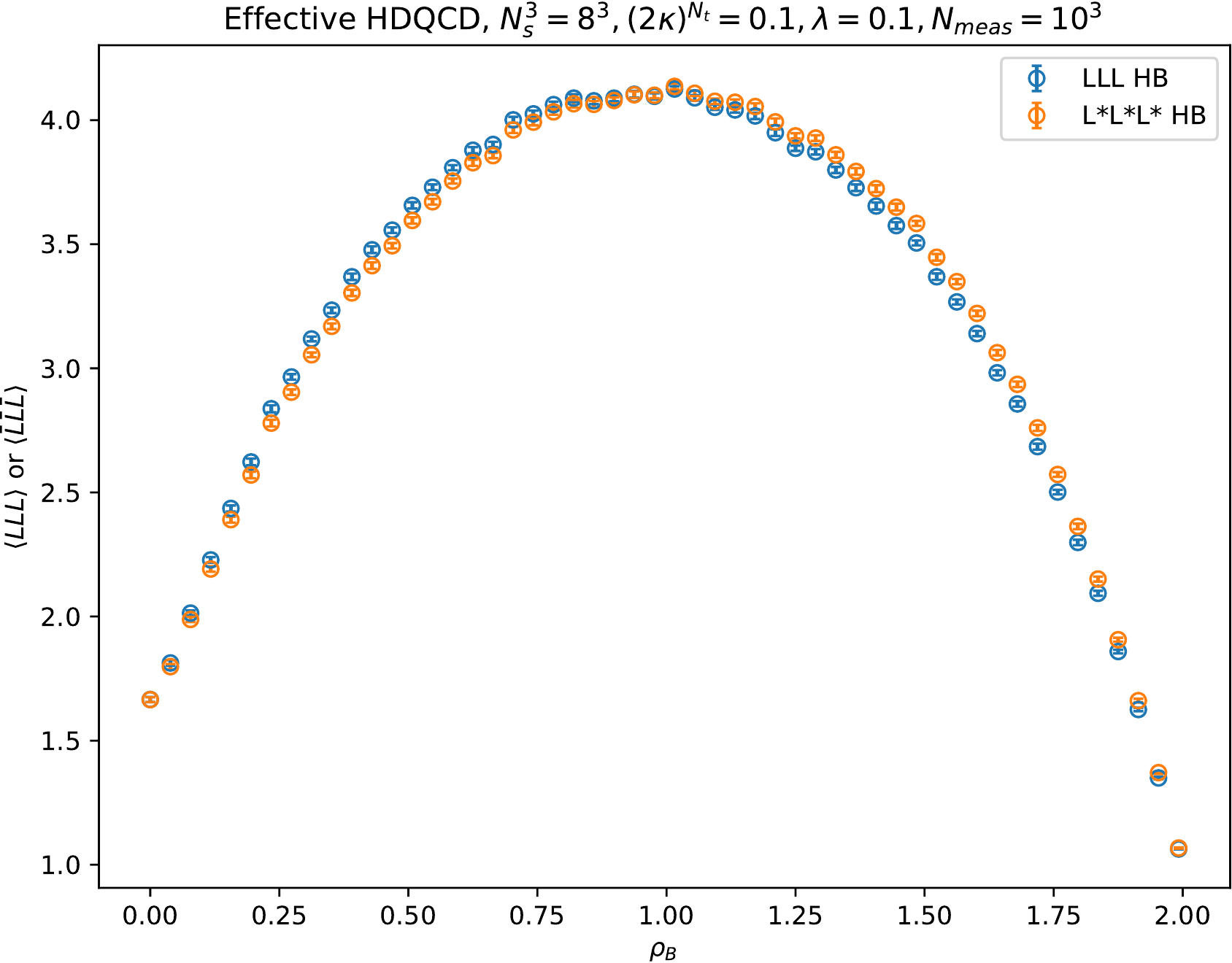} \hfill
\includegraphics[width=0.49\textwidth]{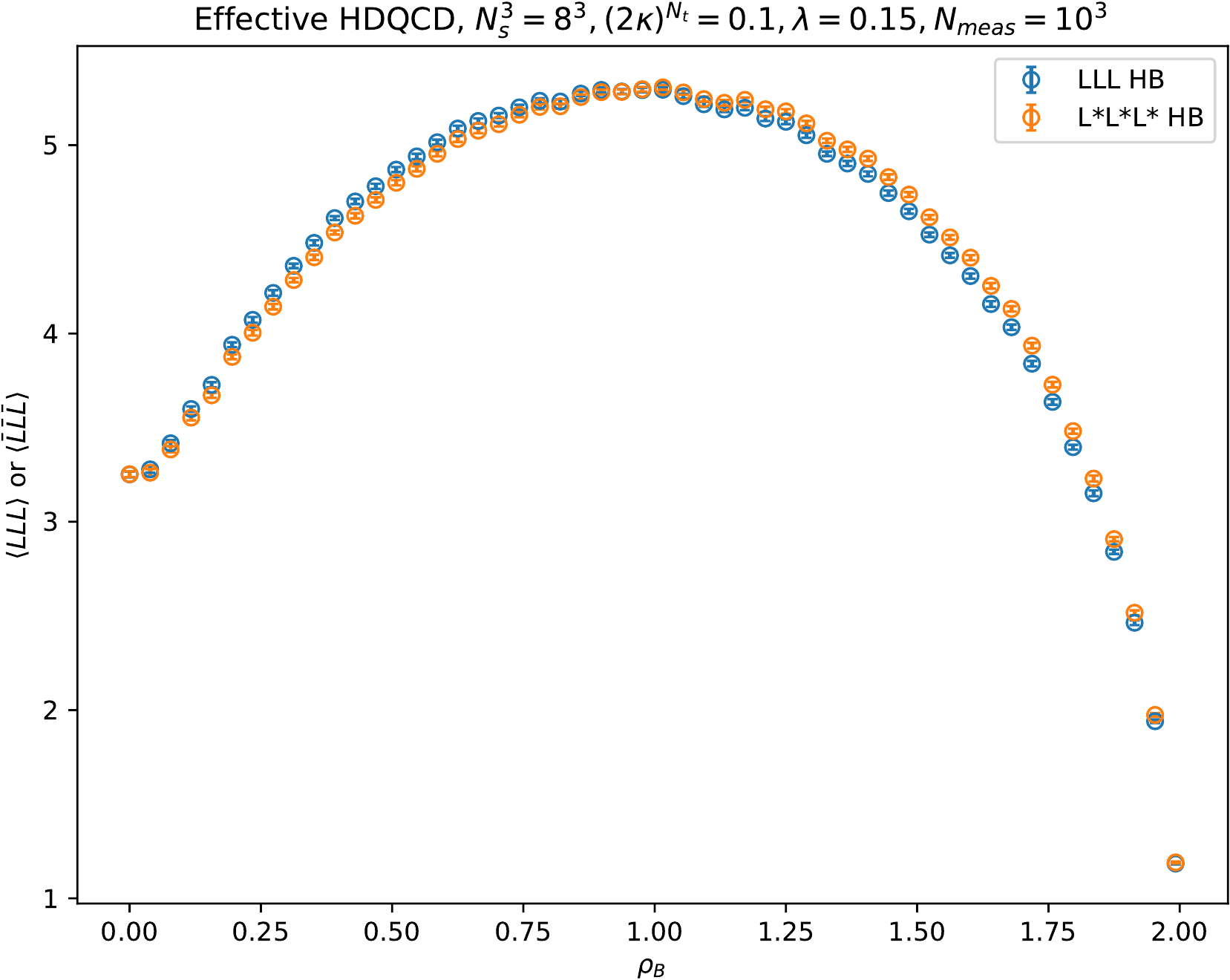}
    \caption{Expectation value of a static baryon or antibaryon in
      the background of finite baryon density for canonical effective
      heavy-dense QCD at two different values of the effective gauge
      coupling, $\lambda=0.1$ ({\it left plot}) and $0.15$ ({\it right
      plot}).\label{fig:(anti-)baryon vs
  rho}}
\end{figure}

\begin{figure}
\includegraphics[width=0.49\textwidth]{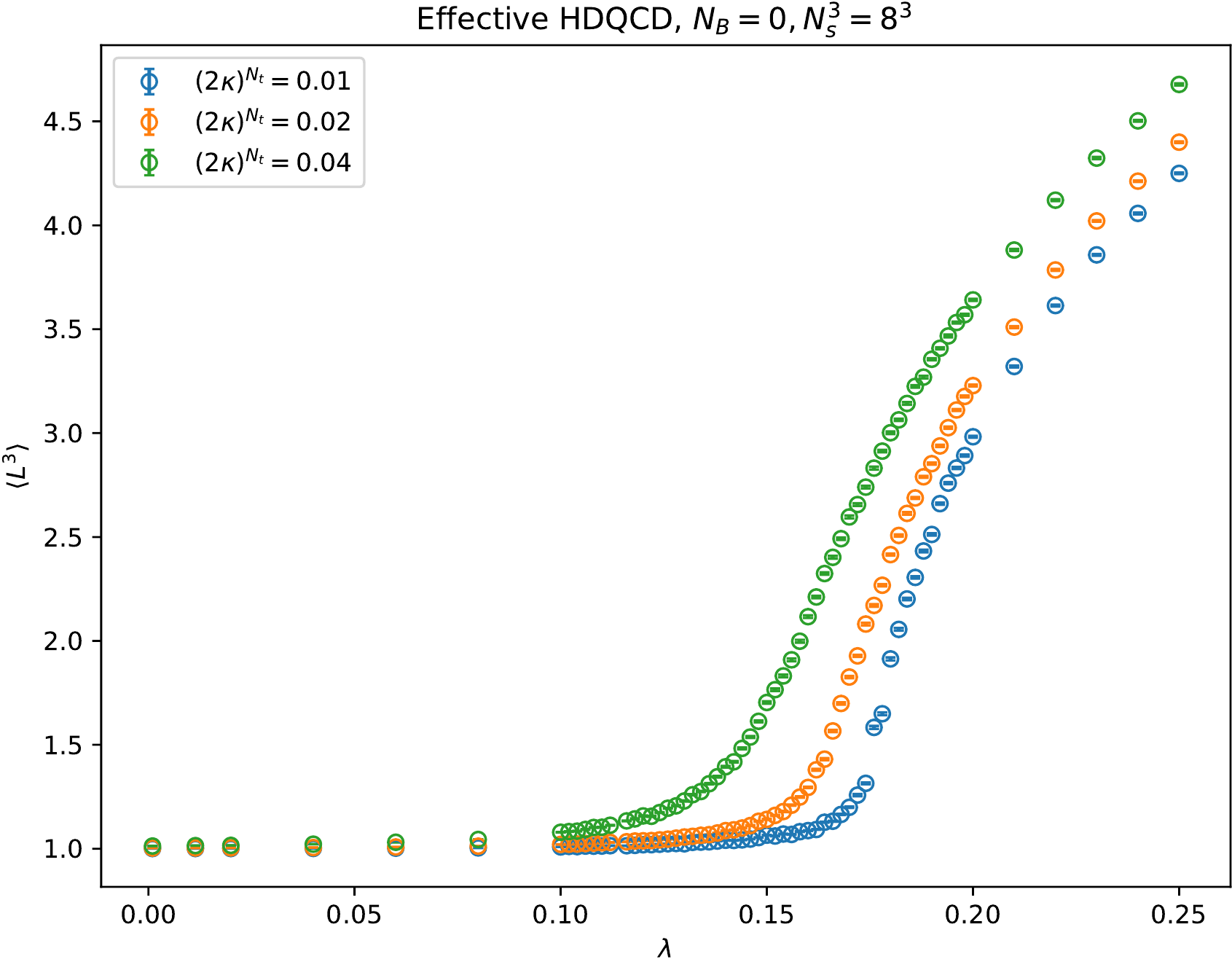} \hfill
\includegraphics[width=0.49\textwidth]{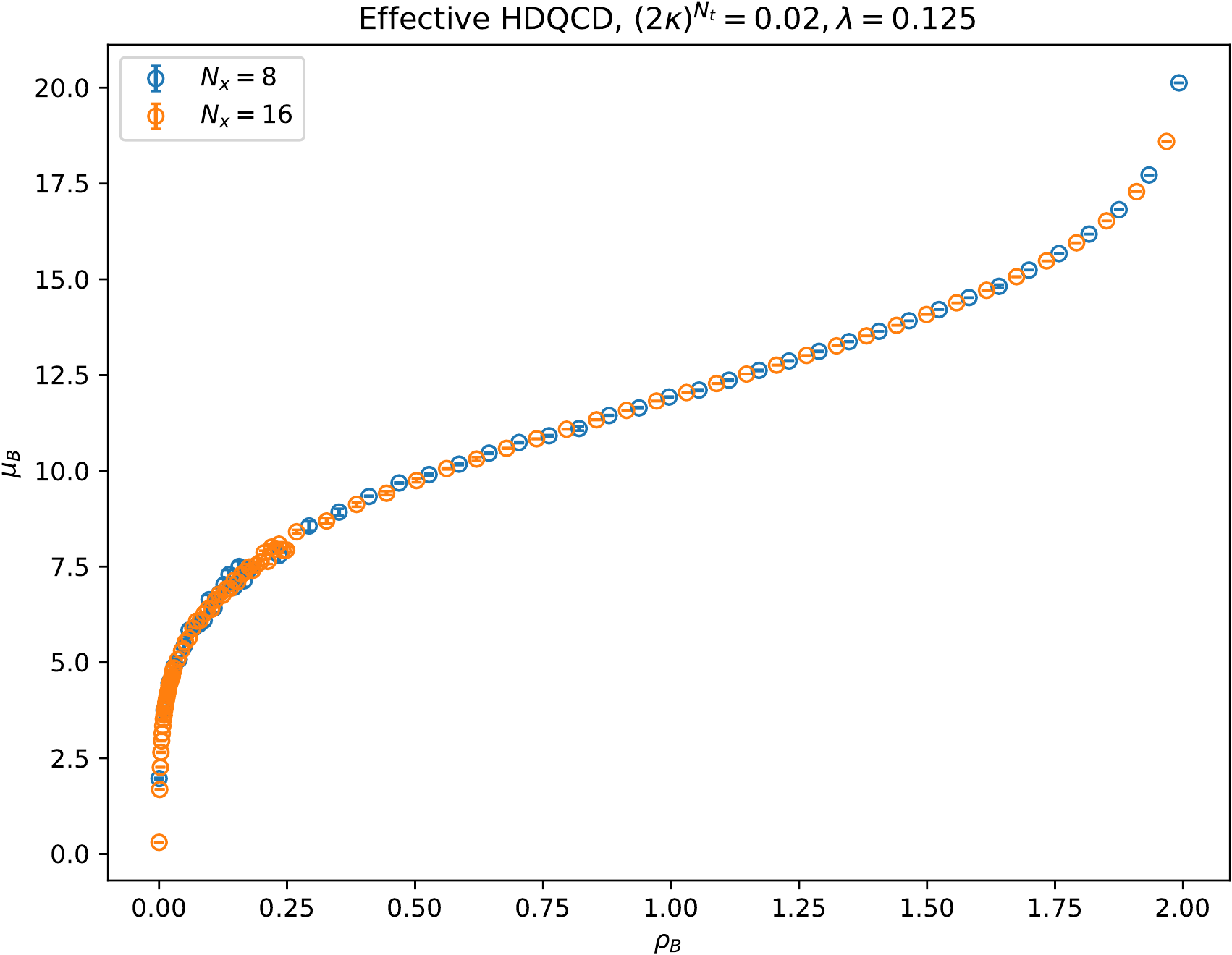}
\caption{{\it Left:} Expectation value of a static baryon at zero
  baryon density as a function of the effective gauge coupling for
  three different temperatures parameterized by
  $(2\kappa)^{N_t}=\exp(-m/T)$. {\it Right:} Baryon chemical potential
  as a function of baryon density at low temperature and for two
  different volumes. \label{fig:baryon vev vs lambda}}
\end{figure}

It is noteworthy that for low densities there is a pronounced change
when going from smaller to larger effective coupling $\lambda$. In the
left plot of Figure \ref{fig:baryon vev vs lambda} we investigate this
in more detail and show the expectation value of a static baryon at
zero density and low temperatures as a function of the effective
coupling $\lambda$. Note that even at zero density, the vacuum is full
of baryon and antibaryons.
The temperature is parameterized by $(2\kappa)^{N_t}=\exp(-m/T)$ and
as we lower it, the transition becomes more pronounced.

Finally, in the right plot of Figure \ref{fig:baryon vev vs lambda} we
show an example of the determination of the baryon chemical potential
as a function of the baryon density. The chemical potential is
calculated from the free energy difference between two systems with
$N_B$ and $N_B+1$ baryons.  As can be seen from comparing the results
at two different volumes, finite size effects are completely
negligible for these parameters. More importantly, the plot
demonstrates nicely how the canonical formulation provides a picture
complementary to the grand-canonical one, in the sense that the
intensive chemical potential parameter is determined as an observable
depending on the extensive number of baryons, in contrast to what is
done in the grand-canonical formulation. As such, the determination of
the chemical potential in the various baryon sectors is equivalent to
a determination of the energies of multi-baryon states.

\section{Summary and outlook}
In these proceedings we reported on the progress in simulating
heavy-dense QCD at fixed baryon number without a sign
problem. Starting from the generic construction of transfer matrices
with fixed fermion numbers applied to the heavy-dense limit of QCD, we
derive an effective Polyakov loop model directly from the underlying
QCD Wilson fermion matrix. In contrast to the grand-canonical
formulation, in the canonical one the $\mathbb{Z}_{3}$ symmetry is
manifest. This in turn leads to a straightforward solution of the sign
problem in the strong coupling limit. After the gauge fields are
integrated out, the only degrees of freedom are configurations of
baryon occupation numbers which need to be summed over. In order to
assess potential advantages of simulations including the gauge field
degrees of freedom in the canonical ensemble compared to the
grand-canonical one, we investigated the anatomy of the sign problem
in both cases. It turns out that the sign problem is less severe in
the canonical formulation by at least an order of magnitude, in
particular at low temperatures the sign problem is so mild that
simulations on very large volumes would be feasible. Finally we
presented some selected results beyond the strong coupling
limit. Using an effective gauge action derived from the usual
plaquette action, the sign problem can also be solved in this case. It
is instructive to see how this is achieved: the effective gauge action
induces fluxes between separated quarks and antiquarks, which binds
them together into clusters. Only clusters with zero triality have
nonvanishing weights which turn out to be positive. Not surprisingly,
this is the same mechanism that is also at work in the Potts model at
fixed fermion number \cite{Alexandru:2010yb}. It also suggests a way
forward to a possible solution for heavy-dense QCD using the standard
plaquette action. While in that case the cluster algorithms are in
general not efficient at temperatures close to the deconfinement
transition, they might be sufficiently efficient at low temperatures
where the constructed clusters remain small.

In the canonical formulation one has explicit
control over the spatial positions of the baryons, which allows us to
calculate, e.g., the static baryon potential, or properties of nuclear
matter along the lines of \cite{deForcrand:2009dh}. Another interesting
application could be the calculation of the energies of multibaryon
states, and of course the determination of the phase diagram as a
function of the effective coupling, the temperature and the baryon density. \\

{\bf Acknowledgements:} We would like to thank Philippe de Forcrand
and Tobias Rindlisbacher for useful discussions.

\bibliographystyle{JHEP}
\bibliography{hdQCDfbnwsp_PoS}

\end{document}